\documentclass[preprint,12pt]{elsarticle}
\usepackage[english]{babel}
\usepackage[onehalfspacing]{setspace}
\usepackage[vmargin={2.5cm,2.5cm},hmargin={2.5cm,2cm},centering]{geometry}
\usepackage{graphicx}

\begin{document}

\title{Damping rates of surface plasmons
for particles of size from nano- to micrometers;
reduction of the nonradiative decay}

\author{K. Kolwas\corref{cor2}}
\ead{Krystyna.Kolwas@ifpan.edu.pl}
\author{A. Derkachova\corref{cor1}}
\ead{Anastasiya.Derkachova@ifpan.edu.pl}

\cortext[cor1]{Corresponding author}
\cortext[cor2]{Principal corresponding author}

\address{Institute of Physics, Polish Academy of Sciences, Al. Lotnik\'{o}w 32/46,
Warszawa, Poland}

\begin{abstract}
Damping rates of multipolar, localized surface plasmons (SP) of gold and silver nanospheres of radii up to $1000nm$ were found with the tools of classical electrodynamics. The significant increase in damping rates followed by noteworthy decrease for larger particles takes place along with substantial red-shift of plasmon resonance frequencies as a function of particle size. We also introduced interface damping into our modeling, which substantially modifies the plasmon damping rates of smaller particles. We demonstrate unexpected reduction of the multipolar SP damping rates in certain size ranges. This effect can be explained by the suppression of the nonradiative decay channel as a result of the lost competition with the radiative channel. We show that experimental dipole damping rates [H. Baida, et al., Nano Lett. 9(10) (2009) 3463, and C. S\"onnichsen, et al., Phys. Rev. Lett. 88 (2002) 077402], and the resulting resonance quality factors can be described in a consistent and straightforward way within our modeling extended to particle sizes still unavailable experimentally.
\end{abstract}

\maketitle


\section{Introduction}

The response of noble-metal nanoparticles to electromagnetic (EM) excitation is dominated by resonant excitations of localized surface plasmons (SPs) (see: \cite{kreibig,KellyCoronadoZhaoSchatz2003,ZhangNogues2008,WilletsVanDuyne2007,Nogues2007} for reviews). When a metal particle is illuminated at corresponding resonance frequency, notably strong surface-confined optical fields can be generated. This property is applied in surface-enhanced Raman
spectroscopy (SERS) \cite{OttoMrozek1992,MockBarbic2002,MaierAtwater},
high-resolution microscopy \cite{FangLeeSunZhang2005} or improvement of
plasmonic solar cells \cite%
{RandPeumansForrest2004,CatchpolePolman2008,PillaiCatchpoleTrupkeGreen2007,NakayamaTanabeAtwater2008,MorfaRowlenReillyRomeroLegematt2008,KimNaJoKimNah2008}. Excitation of SPs at optical frequencies, non-diffraction-limited
guiding of them (e.g. via a linear chain of gold nanospheres \cite%
{MaierAtwater,QuintenLeitnerKrennAussenegg1998,MaierBrongersmaKikAtwater2002,DicksonLyon2000,BrongersmaHartmanAtwater2000,KrennDereuxWeeber1999})
and transferring them back into freely propagating light are processes of great
importance in many applications. Gold and silver nanoparticles are important
components for subwavelength integrated optics and high-sensitivity
biosensors \cite%
{WilletsVanDuyne2007,LiedbergLundstromStenberg1993,AslanLakowiczGedders2005}
due to their chemical inertness and unique optical properties in the visible
to near-infrared spectral range.

Metal nanospheres are the simplest and the most fundamental
structures for studying the basis of plasmon phenomena. Understanding the
resonant interaction of light with plasmonic nanoparticles is also essential
for applications, e.g. for designing useful photonic devices. The
modeling of plasmon properties under the simplest electrostatic
(quasistatic) approximation is valid only for particles much smaller than
the wavelength of light; than size dependence of plasmon properties can be neglected. If a nanosphere is larger, the properties of the
surface plasmons supported by this structure become dependent on the size of the particle \cite%
{kreibig,alkalia,TamChenKunduWangHalas2007,link2000shape,KolwasDerkachovaShopa2009,KolwasDerkachova2010}.
Size dependence of the retarded electronic response of a plasmonic particle is determined by the properties of
metal at the SP resonance wavelength and the presence of the metal-dielectric interface.

Nanoparticles much smaller than optical wavelength first of all exhibit dipolar surface plasmon oscillations. Higher order multipolar resonances
appear as new features in the optical spectra, at frequencies higher than
that of the dipolar plasmon frequency. The spectral signatures of these
higher order plasmon resonances were  observed in elongated gold
and silver nanoparticles \cite%
{KrennSchiderRechberger2000,ShufordRatnerSchatz2005,PayneShufordParkSchatzMirkin2006} first. Fewer experimental investigations directly demonstrated the
multipolar character of the observed resonances in spherical particles \cite%
{KolwasDerkachovaShopa2009,KolwasDemianiukKolwas1997,Sonnichen2002}.

Controlling the spectral properties of a plasmonic nanosphere for
technological or diagnostic applications is not possible without knowing  the direct dependence of plasmon properties on particle size. 	
Still it is  believed (e.g. \cite%
{Nogues2007,link2000shape}) that existing theories do not allow the rigorous direct
calculation of the frequencies of the multipolar SP modes.
However, apart from the plasmon resonance frequencies, also the
plasmon decay rates (damping times) are essential in
controlling the spectral response of the plasmonic particle. Total damping rates of multipolar SPs \cite{KolwasDerkachova2010}
define absorbing and emitting properties of a plasmonic nanoantennas
which can be tuned by particle size. Nanoantenna of the right size can serve as an
effective transitional structure which is able to absorb or to transmit
electromagnetic radiation in plasmonic mechanism. The knowledge and
modeling of plasmon damping effect as a function of particle size are thus
of key interest for the development of plasmonic
nanosystems. The central feature here is the actual enhancement of the
electromagnetic field. It served as a stimulus for us for developing a strict
and direct size characterization of multipolar plasmons resonance
frequencies and plasmon damping rates. Our extended modeling up to the radius $R=1000nm$, and up to SP multipolarity $%
l=10$ allows us to determine the intrinsic properties of plasmonic spheres
in the size range which has never been studied before. We do not
apply any restrictions on the particle radius in relation to the
wavelength of light; our study goes far beyond the particle size
for which the quasistatic approximation is justified (e.g. \cite%
{KellyCoronadoZhaoSchatz2003,Nogues2007}).

The main reason of size sensitivity of SP which we discuss (usually
referred to as the retardation effect (\cite%
{KrahneMorelloFiguerolaGeorgeDekaManna2011} and references therein)) is responsible for important red-shift of the plasmon resonance
frequency for larger nanoparticles. It is accompanied
by the significant increase in damping rate followed by noteworthy decrease
for larger particles. In addition, we discuss the effect called the surface electron scattering effect \cite{kreibig} or interface damping \cite%
{SonnichenFranzlWilkVonPlessen2002} which causes the substantial modification of the plasmon damping rates in particles of sizes comparable or smaller than the mean electron free path. Sometimes called the "intrinsic size effect", it is described by the $1/R$ dependence added to the electron relaxation rate $%
\gamma $\ of the dielectric function \cite%
{kreibig,Nogues2007,LinkElSayed1999}. We perform both types of modeling
including or neglecting the effect of surface scattering, in order to examine
the causes of the size-dependent plasmon features.

Our extended modeling for large particles
reveals new features of the total plasmon damping rate. We demonstrated, for the first time in spherical particles,
the effect of
reducing of multipolar SP damping rates below its low size limit
which is equal to the nonradiative damping rate. Suppression of
the total damping rates proven to exist in certain size ranges allowed a new insight into the role of radiation damping in the plasmon decay mechanism. We also describe changes of the quality Q-factor of SP multipolar resonances with particle size. Q-factor is used for determination of the local field enhancement \cite{SonnichenFranzlWilkVonPlessen2002} and effective susceptibility in nonlinear optical processes \cite{shalaev1996small}.

We also confront the size characteristics resulting from our modeling with
the experimental results obtained for the dipole plasmon for gold
\cite{Sonnichen2002,SonnichenFranzlWilkVonPlessen2002} and silver
\cite{baida2009quantitative} particles of different sizes.
We show that the measured dipole damping rates and quality factors \cite%
{Sonnichen2002,SonnichenFranzlWilkVonPlessen2002} can be described in a consistent
and straightforward way within our modeling which deliver also predictions  for particle of size range still unavailable experimentally.

\section{Size dependence of SP parameters derived from optical spectra}

It is widely accepted, that excitation of surface plasmons (collective
free-electron oscillations) is responsible for the commonly observed
pronounced peaks in the optical absorption or extinction spectra. The
spectra collected for nanoparticles of various sizes are used as a
source of data allowing to reconstruct the change of the resonance (usually dipole
only) position with size (e.g. \cite%
{kreibig,SonnichenFranzlWilkVonPlessen2002,JainHuangAElSayed2007}. The width
of the peak is related to the (dipole) plasmon damping rate \cite%
{Sonnichen2002,SonnichenFranzlWilkVonPlessen2002,baida2009quantitative,hövel1993width,link1999spectral,CharlesGaraAherneLedwith2011}.

Mie theory delivers an indispensable formalism enabling the description
of scattering of a plane monochromatic wave by a homogeneous sphere of known
radius, surround by a homogeneous medium \cite%
{BornWolf,BohrenHuffmann,Stratton,MishenkoTravisLacis2002}. Mie formulas
allow to predict the intensities of light scattered in a given direction
or the absorption and scattering cross-section spectra for particles
\emph{of a chosen size}. The Mie spectra can be compared with
the spectra (intensities of absorbed or scattered light) measured experimentally for particles of \emph{the same  size}.

\begin{figure}[h]
\begin{center}
\doublespacing \scalebox{0.4}{\includegraphics{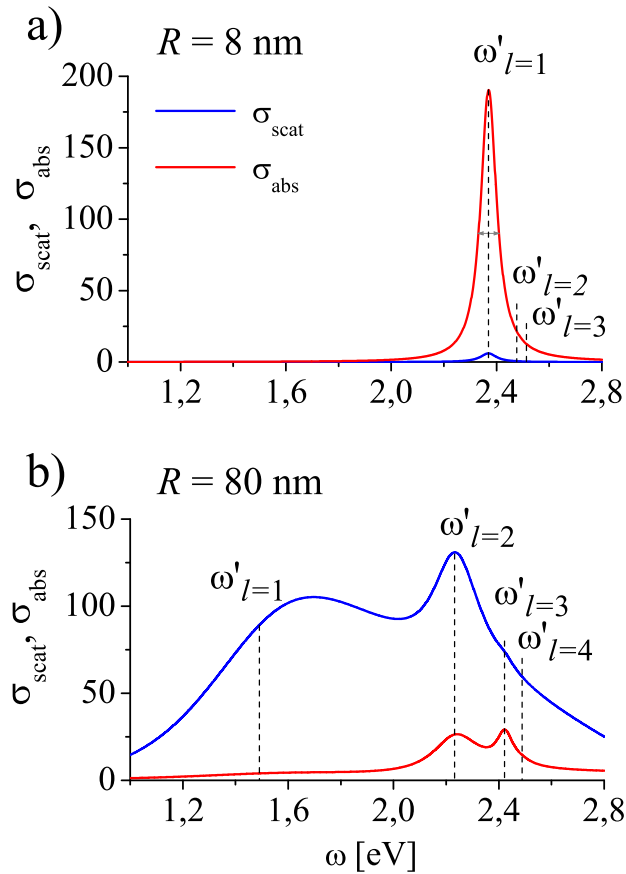}}
\caption{Absorption $C_{abs}(\omega )$ and scattering $C_{scat}(\omega )$ cross-sections for nanospheres of radii a) 8nm and b) 80nm (Mie
theory). Figures illustrate suppression of absorption in larger particles. Broadening, deformation and shift of the maxima in the scattering spectra of larger particles in comparison with the SP resonance eigenfrequencies $\omega_{l}^{\prime }$ is also shown. }\label{Fig_spectra}
\end{center}
\end{figure}

However, predicting \emph{the size dependence} of multipolar plasmon resonance
frequencies, Mie scattering theory is an inconvenient tool. SP
size dependence can be determined only indirectly, by laborious
derivation of positions of maxima in the consecutive spectra collected for various
particle radii. The same applies to derivation of the size dependence of the
(dipole) SP frequency corresponding to the peak position from experimental data.
In addition, the maximum in the spectrum ascribed to SP resonance
can be broadened, deformed and shifted in respect to the frequency of
plasmon oscillation mode \cite{KolwasDerkachovaShopa2009,KolwasDerkachova2010},
as illustrated in Figure \ref{Fig_spectra}. That can affect the plasmon resonance position obtained from the experimental or
calculated scattering spectra. It is even more difficult to determine the SP damping times from the width of the SP resonances, manifested in
the optical spectra  of larger particles, and to understand the derived size dependence then (e.g. \cite{SonnichenFranzlWilkVonPlessen2002,baida2009quantitative,JainHuangAElSayed2007}).
The processes leading to plasmon damping are the subject of extensive research and debate (e.g. \cite{kreibig,SonnichenFranzlWilkVonPlessen2002,hövel1993width,link1999spectral,klar1998surface,heilweil1985nonlinear,liau2001ultrafast}%
. The homogeneous linewidth of the SP resonance was connected to the SP damping time. The properties
of the SP are strongly influenced by this parameter. Mie scattering theory is not sufficient to define a general rule describing the size dependence of SP parameters; the damping rates and the resonance frequencies are not parameters of this theory.

\section{Direct size characterization of multipolar SP modes \label%
{EigenvalueProblem}}

In order to derive parameters of SP as a function of particle size, we use an
accurate electrodynamic approach based on
\cite{rupin}, and described in more details in e.g. \cite{halevi}. The
formalism of Mie theory is used. However, the problem is formulated in
absence of the illuminating light field; we are interested in intrinsic
eigenproperties of the spherical particle (the analogy to the cavity
eigenproblem) \cite{KolwasDerkachovaShopa2009,KolwasDerkachova2010}. In the present paper we extend the numerical calculations up
to the radius of $1000nm$ and plasmon polarity from $l=1$ up to $l=10$.
That allowed us to explore new features of localized SP excitations and reconsider the results of \cite{KolwasDerkachova2010}.

We consider continuity relations at the spherical metal/dielectric interface
for the electromagnetic (EM) fields, which are the solutions of the Helmholtz equation in spherical coordinates ($r,\theta,\phi$) inside and outside the sphere. The transverse magnetic (TM) modes of EM waves localized on
the interface possess a nonzero component of the electric field normal to the surface $E_{r}$, which can couple with
free-electron charge oscillations at the boundary $r=R$. The conditions for the nontrivial solutions of the continuity relations for TM mode define the dispersion relation:

\begin{equation}
\sqrt{\varepsilon _{in}(\omega )}\xi _{l}^{\prime }\left( k_{out}(\omega
)R\right) \psi _{l}\left( k_{in}(\omega )R\right) -\sqrt{\varepsilon
_{out}(\omega )}\xi _{l}\left( k_{out}(\omega )R\right) \psi _{l}^{\prime
}\left( k_{in}(\omega )R\right) =0  \label{DR}
\end{equation}

which is fulfilled for the complex eigenfrequencies of the field $\Omega
_{l} $ in successive multipolar modes $l=1$,$2$,$3$,... . $\psi
_{l}\left( z\right) $ and $\xi _{l}\left( z\right) $ are Riccati-Bessel
spherical functions, the prime symbol ($\prime $) indicates differentiation
with respect to the argument, $k_{in}=\frac{\omega }{c}\sqrt{\varepsilon
_{in}(\omega )}$ and $k_{out}=\frac{\omega }{c}\sqrt{\varepsilon _{out}}$, $%
\varepsilon _{in}(\omega )$ and $\varepsilon _{out}$ are the dielectric
functions of a metal and of the particle dielectric surrounding,
respectively. The dispersion relation (Equation(\ref{DR})) is solved
numerically for complex values of $\Omega _{l}(R)=\omega
_{l}^{\prime }(R)+i\omega _{l}^{\prime \prime }(R)$ ($\omega _{l}^{\prime
\prime }(R)<0$) for $l$ in the range of $1\div 10$. We solved the problem for successive $R$ values from 1 to $1000nm$ in $1nm$ steps. The
\textit{fsolve} function of the MatLab program, utilizing the
\textit{Trust-region dogleg} algorithm was used.

\subsection{Material properties of plasmonic particles\label{MaterialProp}}

The simplest analytic function often used to describe a wavelength dependence of optical
properties of metals like gold or silver \cite%
{EtchegoinRuMeyer2006,LeeElSayed2006,OubreNordlander2004,NoguezRomanVelazquez,OrdalBellAlexanderJr1985}
 results from the Drude-Lorentz-Sommerfeld model:

\begin{equation}
\varepsilon _{D}(\omega )=\varepsilon _{0}-\omega _{p}^{2}/(\omega
^{2}+i\gamma \omega )  \label{epsilon-bulk}
\end{equation}

with $\varepsilon _{0}=1$ (e.g. \cite{kreibig,Nogues2007}). The effective parameter $\varepsilon _{0}>1$
describes contribution of bound electrons, $\omega_{p}$ is the
effective bulk plasma frequency which is associated with effective concentration of
free-electrons, $\gamma$ is the phenomenological damping constant of
electron motion. For bulk metals $\gamma =\gamma _{bulk}$ is
related to the electrical resistivity of the metal and is supposed to
include all microscopic damping processes due to photons, phonons,
impurities and electron-electron interactions. In the present modeling we accept the following effective
parameters of the dielectric function of the bulk gold:\textbf{\ }$%
\varepsilon _{0}=9,84$, $\omega _{p}=9,010eV$, $\gamma _{bulk}=0,072eV$.

The collision time $1/\gamma _{bulk}$ determines the electron mean free path
in bulk metals. At room temperatures the electron mean free path in gold is $
42nm$ \cite{kreibig}. It can be comparable or larger than a dimension of
a particle. An additional relaxation term added to the relaxation rate $%
\gamma $ accounts for the effect of scattering of free electrons by the
surface \cite%
{kreibig,LinkElSayed1999,CharlesGaraAherneLedwith2011,quinten1996optical,NovoGomezPerezJuste2006,HuNovoFunstonWang2008}%
:

\begin{equation}
\gamma _{R}(R)=\gamma _{bulk}+A\frac{v_{F}}{R}  \label{gammaR}
\end{equation}

where $v_{F}$ is the Fermi velocity ($v_{F}=1.4\cdot 10^{-6}m/s$ in gold),
and $A$ is the theory dependent quantity of the order of 1 \cite{kreibig}.
We accept the value $A=0.33$ in our modeling, according to \cite%
{NovoGomezPerezJuste2006}. Then the size-adopted bulk dielectric function is:

\begin{equation}
\varepsilon _{\gamma (R)}(\omega ,R)=\varepsilon _{0}-\omega
_{p}^{2}/(\omega ^{2}+i\gamma _{R}(R)\omega )  \label{epsilonR}
\end{equation}

The real part of $\varepsilon _{\gamma (R)}$ is not modified by surface
scattering. Frequency (and radius) dependence of the dielectric function (Equation \ref{epsilon-bulk} or \ref{epsilonR}) couples to the overall frequency (and radius) dependence of the plasmon dispersion relation (Equation(\ref{DR})).

To underline the crucial role of interface damping in
smaller particles, we will compare the results obtained with the dielectric
function $\varepsilon _{in}=\varepsilon _{D}(\omega )$ given by Equation (%
\ref{epsilon-bulk}) (with $\gamma =\gamma _{bulk}$, surface scattering neglected) and $\varepsilon
_{in}=\varepsilon _{\gamma (R)}(\omega ,R)$ given by Equation (\ref{epsilonR}%
) (with $\gamma =\gamma _{R}(R)$). The dielectric function of the particle
surrounding (vacuum/air) is assumed to be $\varepsilon _{out}=1$, or is
chosen to reflect the index of refraction of the particle environment, as
described in Section \ref{Dipole}, where we compare the results of our
modeling with some experimental results of \cite%
{SonnichenFranzlWilkVonPlessen2002,baida2009quantitative}.

\subsection{Resonance frequencies and damping rates vs radius; the role of the interface damping \label{Solutions}}

The solutions of the dispersion relation (Equation(\ref{DR})) define the
size dependence of multipolar plasmon oscillation frequencies $\omega
_{l}^{\prime }(R)=Re(\Omega _{l}(R))$ and damping rates $|\omega
_{l}^{\prime \prime }(R)|=|Im(\Omega _{l}(R))|$ (and SP damping times $%
T_{l}(R)=\hbar /|\omega _{l}^{\prime \prime }(R)|$) of the EM surface modes
oscillations. We express $\omega _{l}^{\prime }, \omega _{l}^{\prime
\prime },\omega $ and $\gamma $ in electronvolts (eV) for convenience.

\begin{figure}[htb]
\begin{center}
\doublespacing \scalebox{0.4}{\includegraphics{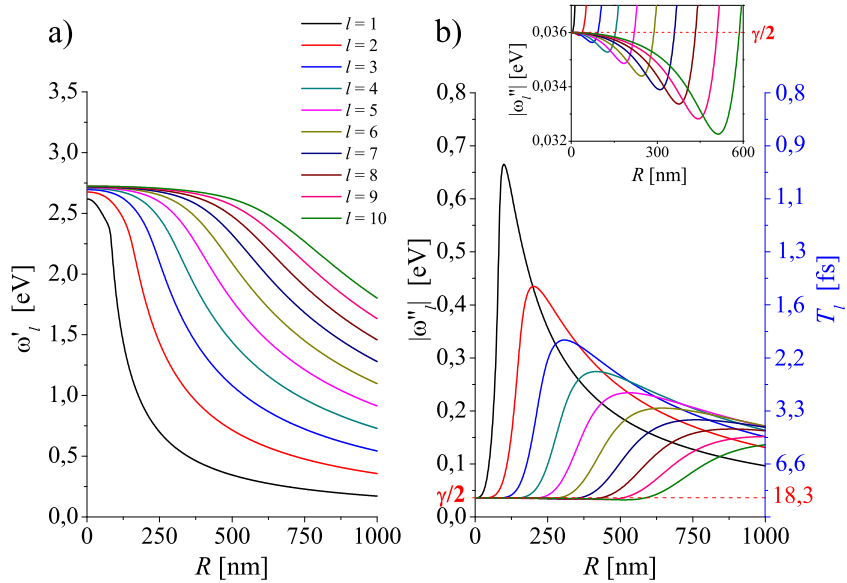}}
\end{center}
\caption{a) Multipolar plasmon resonance
frequencies $\protect\omega _{l}^{\prime }(R)$ and b) plasmon damping
rates $\left\vert \protect\omega _{l}^{\prime \prime }(R)\right\vert $ (left
axis) and corresponding damping times $T_{l}$ (right axis) vs particle radius
$R$. Reduction of the plasmon damping rates $\left\vert \protect\omega %
_{l}^{\prime \prime }(R)\right\vert$ below the nonradiative limit $\gamma/2$)
is demonstrated in the inset.}
\label{Fig_eigenvalues_full}
\end{figure}

The size dependence of $\omega
_{l}^{\prime }(R)$ and $|\omega _{l}^{\prime \prime }(R)|$ of
SP modes for the first ten multipolar plasmons for gold spheres in vacuum/air ($\varepsilon _{out}=1$, $\gamma =\gamma _{bulk}=0.072eV$, the radius up to $1000nm$) is presented in Figure \ref{Fig_eigenvalues_full}.
Black line ($l=1$) represents the dipole resonance frequency. Resonant excitation
of SP oscillations takes place when the frequency of incoming light field $%
\omega $\ approaches the eigenfrequency of a plasmonic nanoantenna of a
given radius $R$: $\omega =\omega _{l}^{\prime }(R)$, $l=1,2,3,...$. If
excited, plasmon oscillations are damped at corresponding rates $|\omega
_{l}^{\prime \prime }(R)|$ (Figure \ref{Fig_eigenvalues_full}b)). Damping of surface plasmon oscillations is the inherent
property of SP modes that defines absorbing and scattering properties of the
plasmonic particle as a function of size.

The SP frequencies $\omega
_{l}^{\prime }(R)$ decrease with size monotonically, as illustrated in Figure \ref%
{Fig_eigenvalues_full}a). For given particle radius $R$, $\omega _{l}^{\prime }(R)$ increase with plasmon polarity $l$.
The multipolar SP resonances for larger particles are better spectrally
resolved than those for the smaller ones.

In the limit of small size, the analytic expressions for $\omega
_{l}^{\prime }(R)$ and $\omega _{l}^{\prime \prime }(R)$ can be found from
the relation (\ref{DR}) after applying the power series expansion of the
spherical Bessel and Hankel functions. Keeping only the first terms of the
power series one can get:

\begin{eqnarray}
\omega _{0,l}^{\prime } &=&\left[ \frac{\omega _{p}^{2}}{\varepsilon _{0}+%
\frac{l+1}{l}\varepsilon _{out}}-\left( \frac{\gamma }{2}\right) ^{2}\right]
^{1/2},\mathbf{\hspace*{0.04in}\ \ \ }  \label{omega0prim} \\
\omega _{0,l}^{\prime \prime } &=&-\frac{\gamma }{2}.  \label{omega0bis}
\end{eqnarray}

For a perfect free-electron metal ($\varepsilon _{0}=1$, $\gamma =0$) and $%
\varepsilon _{out}=1$, Equation (\ref{omega0prim}) leads to the well-known plasmon frequencies within the ''quasistatic approximation'' \cite%
{kreibig,halevi,Boardman}: $\omega _{0,l}^{\prime }=\omega _{p}/\sqrt{%
1+\varepsilon _{out}(l+1)/l}$, and in particular to the giant Mie resonance
frequency $\omega _{0,l=1}^{\prime }=\omega _{p}/\sqrt{3}$ for $l=1$.

In the smallest particles, the plasmon damping rates $|\omega _{l}^{\prime
\prime }(R)|$ fall to pure nonradiative damping rates $|\omega
_{0,l}^{\prime \prime }|=\gamma ^{nrad}=\gamma /2$,
as demonstrated in Figure \ref{Fig_eigenvalues_full}b). The values of $|\omega
_{0,l}^{\prime \prime }|$ are the same for all multipolar plasmon
modes $l=1,2,3...$ (Equation (\ref{omega0bis})). The nonradiative damping
rates $\gamma ^{nrad}$ are due to the ohmic losses if interface damping is
neglected ($\gamma =\gamma _{bulk}$), or are equal to $\gamma_{R}/2$ (Equation \ref{gammaR}) if interface damping is included in the
modeling.

\begin{figure}[h]
\begin{center}
\doublespacing \scalebox{0.4}{\includegraphics{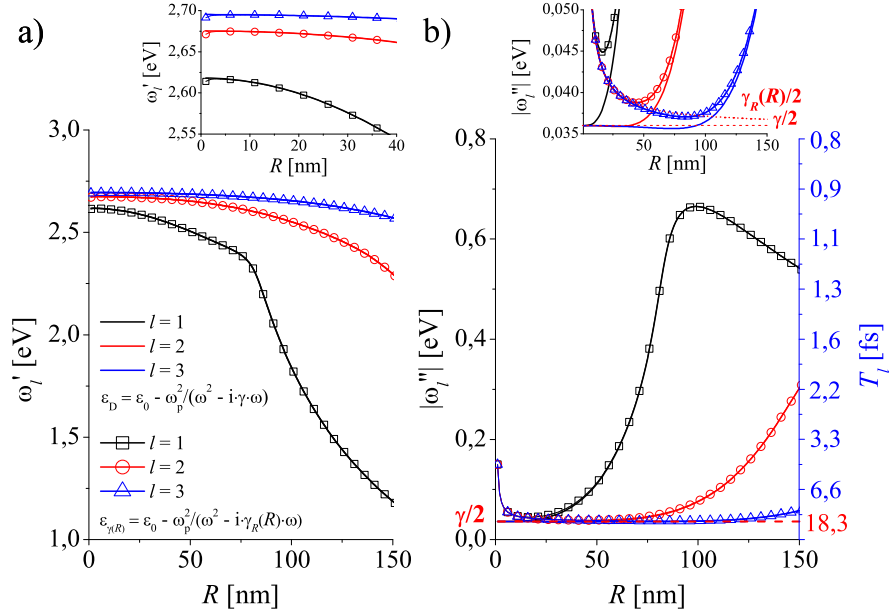}}
\end{center}
\caption{a) Dipole, quadrupole and hexapole plasmon resonance frequencies $%
\protect\omega_{l}^{\prime }(R)$ and b) corresponding plasmon damping rates $%
\left| \protect\omega_{l}^{\prime \prime }(R)\right| $ calculated without
(solid lines) and with (lines with hollow
circles) interface damping taken into account. Insets are magnifications of figures a) and b)
for particles of small radii.}
\label{Fig_eigenvalues_3l}
\end{figure}

Interface damping practically does not influence the size dependence of $%
\omega _{l}^{\prime }(R)$ as illustrated in Figure \ref{Fig_eigenvalues_3l}%
a). The minute red shift for the smallest sizes due to interface
damping is shown in inset of Figure \ref{Fig_eigenvalues_3l}a). A
similar finding results from the experimental observations \cite%
{Sonnichen2002,SonnichenFranzlWilkVonPlessen2002}. Therefore, the red shift
of the (dipole) SP resonance frequency sometimes observed in small particles
versus decreasing size must result from some other phenomena due to the
complex chemical effects and uncertainties in experimental samples \cite%
{PengMcMahonSchatzGraySun2010}.

In extended size range that we study, plasmon damping rates $|\omega _{l}^{\prime \prime }(R)|$ are not simple monotonic functions of the radius
(Figure \ref{Fig_eigenvalues_full}b)). The initial fast
increase of $|\omega _{l}^{\prime \prime }(R)|$
with size is followed by gradual decrease for sufficiently large radii $R$.
The increase in $|\omega _{l+1}^{\prime \prime }(R)|$ in the subsequent $l+1$ SP
modes is followed by the decrease in the SP damping rates $|\omega
_{l}^{\prime \prime }(R)|$ of lower polarity modes.

\begin{figure}[h]
\begin{center}
\doublespacing \scalebox{0.4}{\includegraphics{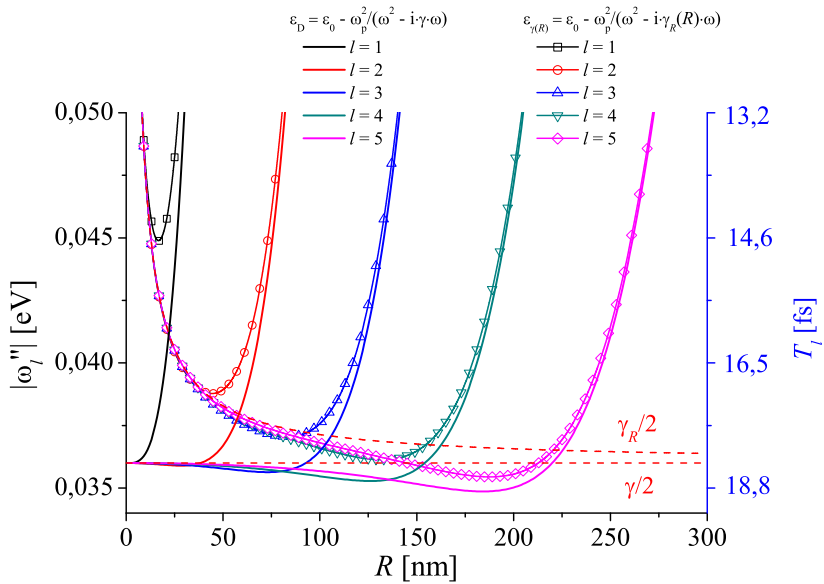}}
\end{center}
\caption{Demonstration of the reduction effect of plasmon damping
rates $\left| \protect\omega_{l}^{\prime \prime }(R)\right|$
calculated without ($\protect\gamma =\protect\gamma _{bulk}$,
solid lines) and with ($\protect\gamma =\protect\gamma _{R}(R)$,
lines with hollow circles) interface damping; the magnification of a part
of Figure \protect\ref{Fig_eigenvalues_3l}.}
\label{Fig_OmegaBisSmall}
\end{figure}

The surface scattering effect affects the total damping rate $%
|\omega _{l}^{\prime \prime }(R)|$, as illustrated in Figure \ref%
{Fig_eigenvalues_3l}b) and in the magnification of the part of this graph
presented in Figure \ref{Fig_OmegaBisSmall}. The surface scattering
contribution to the total relaxation rate $\gamma _{R}(R)$
looses its importance for particles of radius larger than $\sim
300nm$ (using the criterion $(\gamma _{R}-\gamma )/\gamma _{R}\approx
1\% $).

\section{Radiative and nonradiative contributions to the total SP damping
rates \label{Contributions}}

The processes leading to damping of SP have been the subject of wide
discussion and extensive studies, e.g.
\cite{kreibig,Sonnichen2002,SonnichenFranzlWilkVonPlessen2002,HeilweilHochstrasser1985,LamprechtKrennLeitnerAussenegg1999,SteitzBosbachWenzelVartanyan2000}
which, however, were limited to the dipole case. It was accepted that
(dipole) SP in metal nanoparticles decays through both inelastic processes
and elastic dephasing process which were usually neglected. Inelastic
processes can be further divided into radiative and non-radiative decay
processes \cite%
{KrahneMorelloFiguerolaGeorgeDekaManna2011,SonnichenFranzlWilkVonPlessen2002,HeilweilHochstrasser1985}%
. The SP damping time $T$ was determined from the homogeneous
linewidth $\Gamma =2\hbar /T$ of a
maximum in the spectrum due to the SP dipole resonance. The observed spectral broadening of the maxima with size due to radiation is sometimes expected to be proportional to the particle volume \cite%
{SonnichenFranzlWilkVonPlessen2002,NovoGomezPerezJuste2006,melikyan2004surface}%
. Our results show (Figure \ref{Fig_eigenvalues_full}b)), that the
multipolar plasmon damping rates $|\omega _{l}^{\prime \prime }(R)|$ are not
simple, monotonic functions of particle radius. The initial fast increase
of the total SP damping rate $|\omega _{l}^{\prime \prime }(R)|$ with size
is followed by gradual decrease for sufficiently large radii for the given mode $l
$. Our modeling suggest that the proportionality of the multipolar radiation
damping rate to $R^{3}$ is not a general rule (see Figure %
\ref{Fig_eigenvalues_full}b)).

Our extended modeling allows us to consider the higher order multipolar
damping rates $\left\vert \omega _{l}^{\prime \prime }(R)\right\vert$
l=1,2...10, as well. The total multipolar damping times $\hbar /T_{l}$ can be
related to the corresponding damping rates (homogeneous linewidths of the absorption
\cite{KolwasDerkachova2010}\ spectra $\Gamma _{l}(R)$) and their
size dependencies in a natural way:

\begin{equation}
\Gamma _{l}(R)=2\left\vert \omega _{l}^{\prime \prime }(R)\right\vert
=2\hbar /T_{l}
\end{equation}

and

\begin{equation}
\left\vert \omega _{l}^{\prime \prime }\right\vert =\gamma _{l}^{rad}+\gamma
^{nrad}  \label{gamma_rad and nrad}
\end{equation}

where $\gamma ^{nrad}=\gamma /2$. The assumption, that the nonradiative and
radiative processes (in dipole plasmon mode) are additive and are
independent, is commonly accepted (see for example: \cite%
{KolwasDerkachova2010,CharlesGaraAherneLedwith2011,NovoGomezPerezJuste2006,HuNovoFunstonWang2008,HeilweilHochstrasser1985}%
). However, expectation that the size dependence of the total damping rate $%
\left\vert \omega _{l}^{\prime \prime }(R)\right\vert $ is due to the size
dependence of the radiative damping rate $\gamma _{l}^{rad}(R)$ only ($%
\left\vert \omega _{l}^{\prime \prime }(R)\right\vert =\gamma
_{l}^{rad}(R)+\gamma ^{nrad}$, \cite{KolwasDerkachova2010}) must be
reconsidered (see Section \ref{Suppression} bellow).

If the surface scattering effect is included, $|\omega _{l}^{\prime \prime }(R)|$
becomes a decreasing function of size,
starting from smaller particles (see Figure \ref{Fig_eigenvalues_3l}b) and %
\ref{Fig_OmegaBisSmall}, lines with symbols). The total SP damping rates follow the size dependence of
nonradiative damping: $\gamma _{R}(R)/2$: $|\omega _{l}^{\prime \prime
}(R)|\approx \gamma _{R}(R)/2=\gamma ^{nrad}(R)$ in the range of radii which
extends to larger $R$ for growing SP polarity $l$. After reaching the
minimum, $|\omega _{l}^{\prime \prime }(R)|$ tends to follow the radius dependence
unaffected by the interface damping.

A contribution of the radiation damping $\gamma _{l}^{rad}(R)$
to the total damping rate $|\omega _{l}^{\prime \prime }(R)|$\ can be
treated as a measure of the ability of the particle to couple to
the incoming field and to emit light in plasmonic mechanism. As long
as $|\omega _{l}^{\prime \prime }(R)|\simeq \gamma ^{nrad}$, the SP mode $l$
can only weakly couple with the incoming radiation and has weak
radiative abilities. Consequently, the weakly radiative plasmons appear
in the absorption spectra with a smaller amplitude and in scattering
 spectra (see Figure \ref{Fig_spectra}) they hardly manifest. With increasing $l$, SP plasmons gain the radiative character starting from
larger particles due to the fact that
$\gamma _{l>1}^{rad}(R)>\gamma _{l-1}^{rad}(R)$. Therefore, the
higher order SP modes can be excited (and appear in the scattering
spectra) for larger particles. This fact is known from Mie scattering
theory, but its physical sense have not been explained by Mie solutions.

\section{Effect of reduction of multipolar plasmon damping rates \label%
{Suppression}}

In the easier case (interface damping omitted in the modeling), reduction of
the total SP dumping rates $\gamma ^{nrad}$ bellow the $\gamma/2$ value (see the inset in Figure \ref{Fig_eigenvalues_full}b)) manifests for plasmon modes with $l>1$. This reduction effect can be clearly distinguished from other size dependent
processes in some size ranges which grow with increasing $l$; for the
quadrupole ($l=2$) plasmon mode, the reduction effect extends up to $R\simeq 40nm$, for the hexapole ($l=3$) plasmon mode up to $R\simeq $ $92nm$, for $%
l=4$ up to $R\simeq $ $155nm$ and so on.

We have checked, the reduction of the total damping rate is not present if $\gamma=0$ (\ref{epsilon-bulk}). On the contrary, it does not disappear if $\gamma\neq0$ and $\varepsilon _{0}=1$. That suggests, that if the additive form of the Expression (\ref%
{gamma_rad and nrad}) holds, reduction of the total plasmon damping rate is due to the decrease
of nonradiative decay $\gamma ^{nrad}$ as compared with its low size limit $%
\gamma ^{nrad}=\gamma/2$ resulting from the energy dissipation of meatal (absorption). Consequently, the radiative and nonradiative
contributions have to be coupled by their size dependence. The model with interface damping included, reproduces the SP rate reduction
below $\gamma ^{nrad}=\gamma_{R}/2$ (see Figure \ref%
{Fig_OmegaBisSmall}), as well. \bigskip

We can conclude, that reduction of $\left\vert\omega _{l}^{\prime \prime}(R)\right\vert$ below the $\gamma/2=\gamma _{bulk}/2$
(or $\gamma/2=\gamma _{R}/2$) value takes place in the regions of sizes where the nonradiative
damping is still not dominated by the fast radiative damping. Reduction in
the total plasmon damping rate must be connected with suppression of the nonradiative decay
channel. It can be inferred then, that the radiative and nonradiative processes
are not independent. The effective rate of the nonradiative damping must be
size dependent: $\gamma ^{nrad}=\gamma ^{nrad}(R)$ with the value for the
small particle limit: $\gamma _{0}^{nrad}=\gamma /2.$ Suppression of
the nonradiative damping is not restricted to the size ranges for which the effect
was demonstrated, but influences the optical
properties of plasmonic particles in a large range of sizes.
Reduction of $\gamma ^{nrad}(R)$ manifests in the absorption spectrum of particles; absorptive abilities of large particles are poor as compared with small particles. This fact is described by solutions of Mie scattering theory,
but its physical meaning have not been
explained. \bigskip

The effect of the total damping rate reduction with particle dimension was observed for the first
time in the experiment with gold nanorods \cite%
{SonnichenFranzlWilkVonPlessen2002}. The size dependence of the dipole
plasmon damping rate was deduced from the homogeneous linewidth $\Gamma
=2\hbar /T$ of the maxima in the scattered intensities both for nanorods
with various aspect ratios and spherical nanoparticles with radii in
the range from $10$ to $75nm$. For nanorods, (dipole) plasmon damping
rates decreased significantly for lower dipole plasmon oscillation
frequencies. However, the decrease in the damping rate was not  ascribed to the radiative processes.
The authors of explain this effect as reduction nonradiative plasmon decay by the fact that interband excitations in gold require a threshold energy
of about $1.8eV$  and expect, that suppression of the
damping rate in such mechanism would also be present in gold spheres, but for
plasmon resonance energies below $1.8eV$. Suppression of the plasmon damping rates in spheres had not been observed
experimentally, as far as we know; the conclusive experimental data are limited to the plasmon dipole only, while according to our study the effect manifests for $l>1$. Importance of the threshold energy at $1,8eV$ in gold and exclusion of
the radiative damping as a reducing mechanism was not confirmed by our study. On the contrary,
the reduction of multipolar SP nonradiative damping rates results from
competition between radiative damping $\gamma _{l}^{rad}(R)$ and
all other damping processes included in $\gamma ^{nrad}(R)$ and is present in plasmonic particles of any material.

\section{Quality factor of multipolar plasmon resonances \label%
{Quality_factor}}

Enhancement of optical response in resonance is usually described by the
quality factor $Q$, defined as the product of the resonance center frequency
and the bandwidth. In case of plasmonic particles the quality factor is
interpreted as a measure of the local field enhancement
\cite{SonnichenFranzlWilkVonPlessen2002} and is expected to define the effective
susceptibility in nonlinear optical processes \cite{shalaev1996small}.

\begin{figure}[h]
\begin{center}
\doublespacing \scalebox{0.4}{\includegraphics{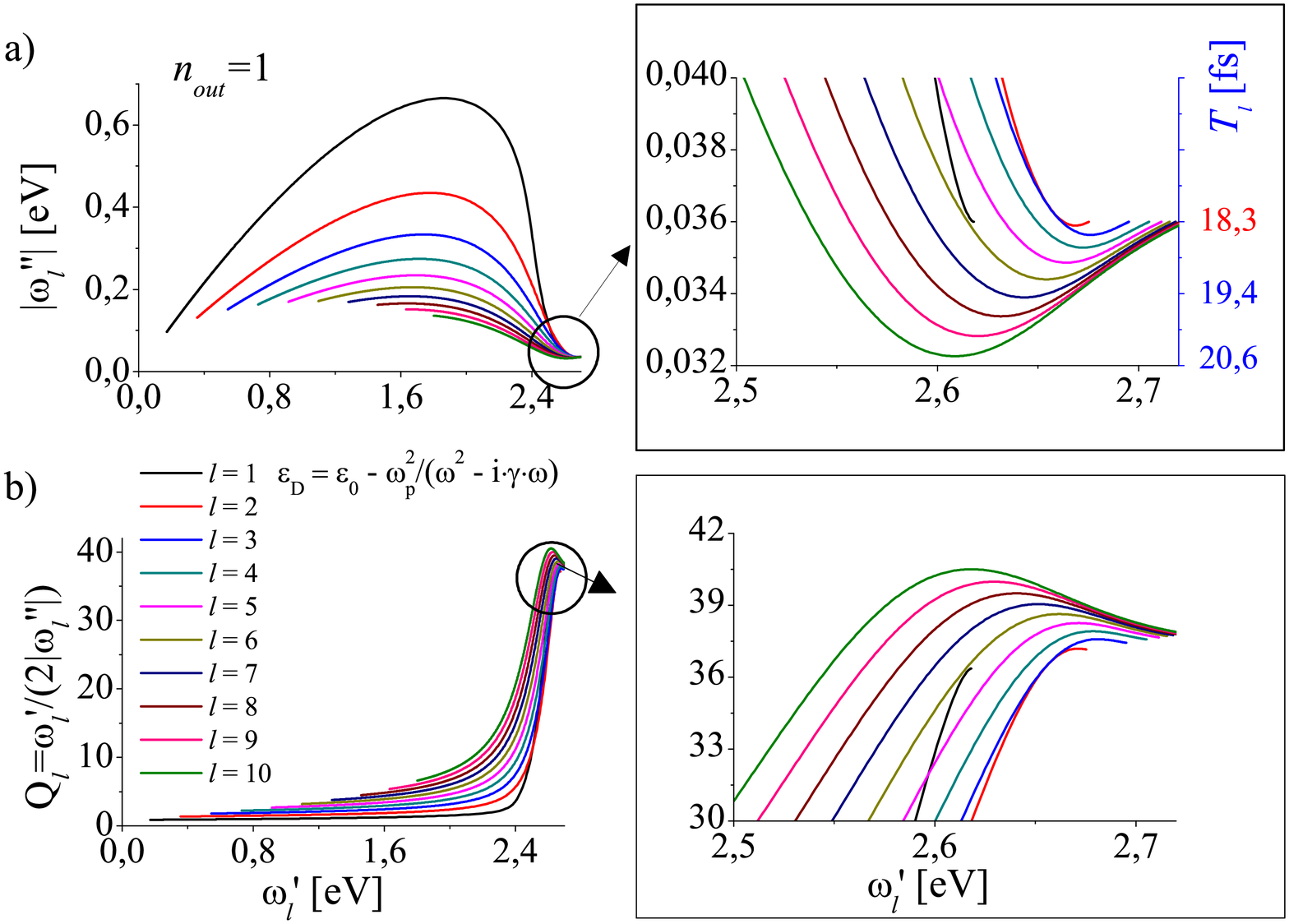}}
\end{center}
\caption{a) Damping rates $|\protect\omega _{l}^{\prime \prime}|$ vs $%
\protect\omega _{l}^{\prime }$ for successive radii $R$ and b) quality factors
$Q_{l}=\protect\omega _{l}^{\prime }(R)/(2\left\vert \protect%
\omega _{l}^{\prime \prime }(R)\right\vert )$ for multipolar plasmon modes
of gold nanoparticles ($n_{out}=1$, surface scattering neglected).
Graphs on the right are magnifications of circled regions of the graphs on the left.}
\label{Fig_QodOega}
\end{figure}

\begin{figure}[h]
\begin{center}
\doublespacing \scalebox{0.4}{\includegraphics{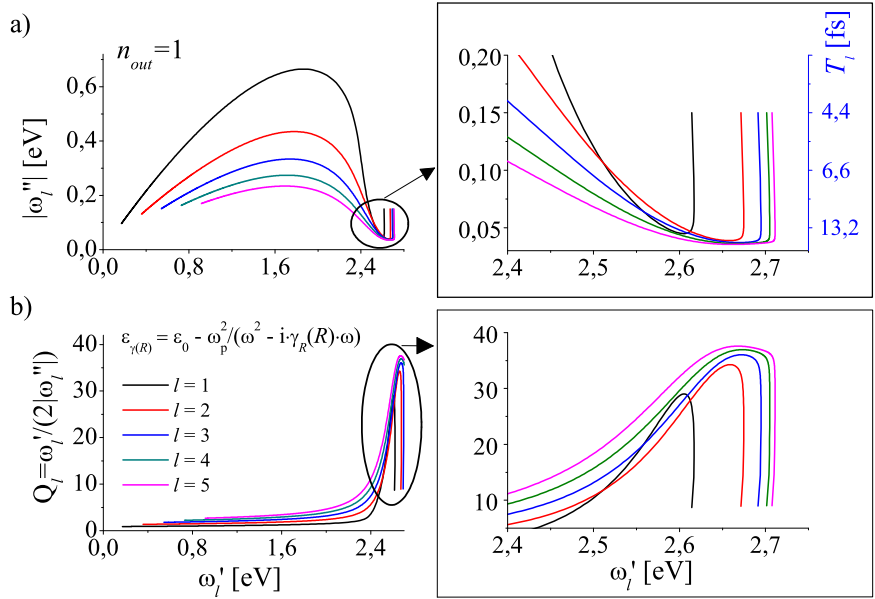}}
\end{center}
\caption{a) Damping rates $|\protect\omega _{l}^{\prime \prime}|$ vs $%
\protect\omega _{l}^{\prime }$ for successive radii $R$ and b) quality factors
$Q_{l}=\protect\omega _{l}^{\prime }(R)/(2\left\vert \protect%
\omega _{l}^{\prime \prime }(R)\right\vert )$ for multipolar plasmon modes
of gold nanoparticles ($n_{out}=1$, surface scattering included).
Graphs on the right are magnifications of circled regions of the graphs on the left.}
\label{Fig_QodOega_SurfScat}
\end{figure}

The dependence of the $\omega _{l}^{\prime \prime }$ vs $\omega _{l}^{\prime
}$ for successive radii $R$ and the resulting quality factors $%
Q_{l}(R)=\omega _{l}^{\prime }(R)/(2\left\vert \omega _{l}^{\prime \prime
}(R)\right\vert )$ for multipolar plasmon modes of gold nanoparticles ($%
n_{out}=1$) are presented in Figures \ref{Fig_QodOega} (surface
scattering neglected) and \ref{Fig_QodOega_SurfScat} (surface
scattering included). The graphs on the right are the magnifications
of the graphes (circled regions) on the left.

\begin{figure}[h]
\begin{center}
\doublespacing \scalebox{0.4}{\includegraphics{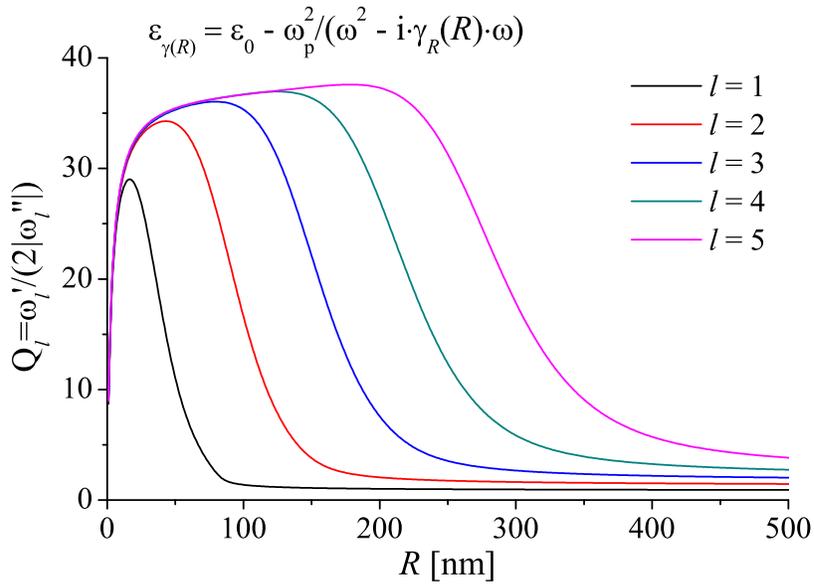}}
\end{center}
\caption{Quality factors $Q_{l}$ vs radius $R$ for multipolar plasmon modes
of gold nanoparticles ($n_{out}=1$, surface scattering included).}
\label{Fig_Ql_odR_surfscat}
\end{figure}

Figure \ref{Fig_QodOega}a) illustrates the effect of the total plasmon
damping rate reduction to the value bellow $\gamma ^{nrad}=0,032eV$ for modes
with $l>1$. Figure \ref{Fig_QodOega}b) illustrates
the resulting quality factors $Q_{l}$ for successive SP modes which we use
as a measure of energy stored in the SP oscillation of plasmon mode $l$
at the resonance frequency $\omega _{l}^{\prime }(R)$. In some ranges of SP resonance frequencies,
the higher polarity plasmon modes ($l>1$) are more efficient in storing the
SP oscillation energy. The similar conclusion comes from Figure \ref%
{Fig_QodOega_SurfScat}b) (interface damping included). Quality
factor of the dipole plasmon resonance reaches maximum value $%
Q_{l=1}\approx 29$ for $\omega _{l}^{\prime
}(R)\approx 2,606eV$ for gold nanoparticle of the radius $R\approx 16nm$
(see Figure \ref{Fig_Ql_odR_surfscat}) and is the fast decreasing function of size for both smaller and larger
particles. If a gold particle is embedded in a medium of higher optical
density (see Figure \ref{Fig_exp_gold}, $n_{out}=1,5$), the maximum quality
factor of the dipole plasmon resonance is even smaller, $Q_{l=1}\approx 26$. Such value corresponds to the resonance frequency of a
gold particle of radius $R\approx 11nm$.

\section{Dipole plasmon damping rates vs size: comparison with
experimental results for silver and gold nanoparticles \protect\cite%
{SonnichenFranzlWilkVonPlessen2002,baida2009quantitative}\label{Dipole}}

\begin{figure}[h]
\begin{center}
\doublespacing \scalebox{0.4}{\includegraphics{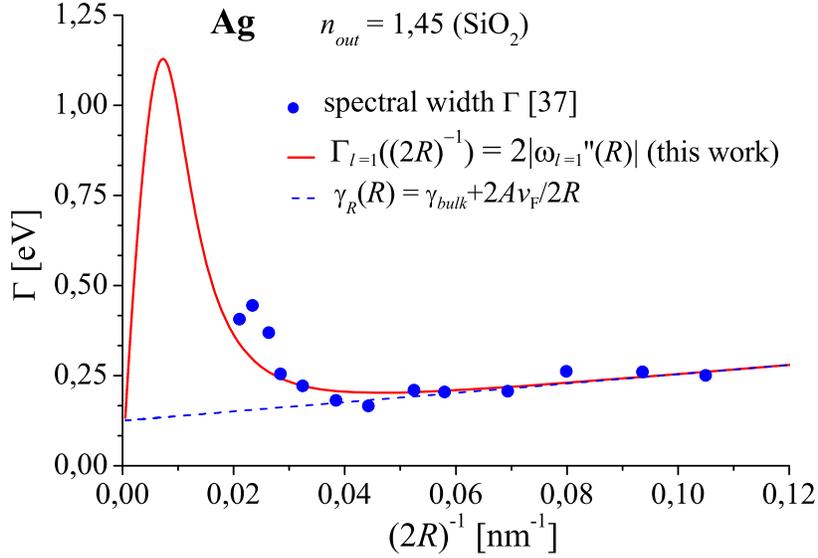}}
\end{center}
\caption{Spectral width $\Gamma $ of plasmon resonances measured \protect\cite{baida2009quantitative} in single spherical Ag@SiO$_{2}$ particles
vs the inverse of their equivalent diameter. Dashed
line represents $\protect\gamma _{R}((2R)^{-1})$ dependence (equation
(\ref{gammaR})), with the parameters $A=0.7$ and $\protect\gamma %
_{bulk}=0.125eV$ resulting from linear fit
the line to the experimental data \protect\cite{baida2009quantitative}. Solid line results from our model with
the parameters $\protect\omega _{p}=9.10,\protect\varepsilon _{0}=4eV,$
$\protect\gamma _{bulk}=0.125eV$ and $A=0.7$.}
\label{Fig_exp_silver}
\end{figure}

\begin{figure}[h]
\begin{center}
\doublespacing \scalebox{0.35}{\includegraphics{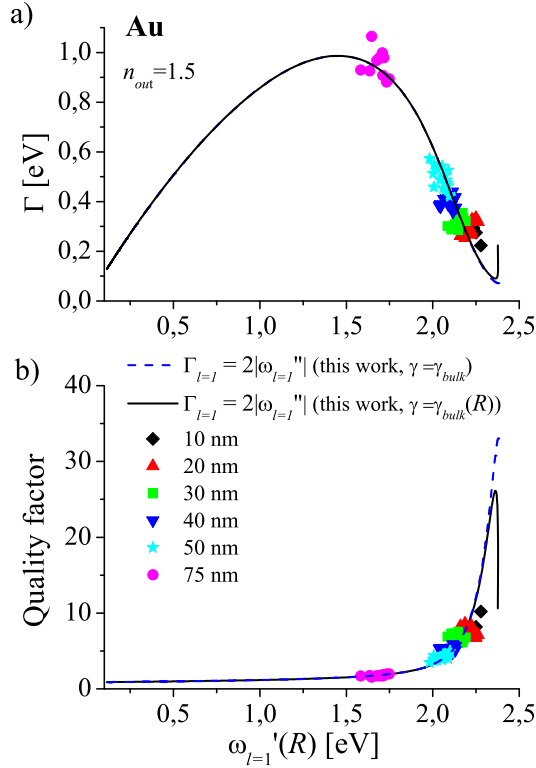}}
\end{center}
\caption{a) Spectral widths $\Gamma$
of plasmon resonances vs resonance energy measured
\cite{SonnichenFranzlWilkVonPlessen2002} in single spherical Au particles.
b) Quality factor resulting from the experimental data.
Lines represent the spectral width $\Gamma_{l=1}(\protect\omega %
_{l=1}(R))=2|\protect\omega _{l=1}^{\prime \prime}(R)|$ and the quality
factor $Q_{l=1}=\protect\omega _{l=1}^{\prime}(R)/2|\protect\omega _{l=1}^{\prime
\prime}(R)|$ vs $\protect\omega _{l=1}^{\prime}(R)$ for the dipole mode
obtained from our modeling with surface damping neglected (dashed line)
and included (solid line).}
\label{Fig_exp_gold}
\end{figure}

Experimental investigations of spectral properties of plasmonic particles
are usually performed on large particle ensembles, where inevitable
variations in size, shape and surface properties tend to mask the spectral
properties of the individual particles. The effects of surface chemistry
(especially in Ag particles) and uncertainties in sample parameters affect linewidths and maxima positions which define damping times and resonance frequencies of SP. We have chosen two experiments
\cite{SonnichenFranzlWilkVonPlessen2002,baida2009quantitative}, which provide the
spectroscopic data for individual spherical particles of silver and gold. The
experiments were performed in well-controlled particle environments as
a function of size for nanoparticles up to $R=25nm$ for silica-coated silver (Ag@SiO$_{2}$)
and for gold nanoparticles immersed in
a index matching fluid ($n_{out}=1.5$) for relatively large range of radii (up to $R=75nm$). However,
well resolved data was reported only for the dipole plasmon resonance.
In Figures \ref{Fig_exp_silver} and \ref{Fig_exp_gold} we compare the
experimental results presented in
\cite{SonnichenFranzlWilkVonPlessen2002,baida2009quantitative} with
the results of our modeling.

Figure \ref{Fig_exp_silver} illustrates the spectral width $\Gamma$ measured in
different single Ag@SiO$_{2}$ nanoparticles vs the inverse of
their equivalent diameter $D_{eq}$, optically determined by fitting the
extinction spectra (see Figure 4 in \cite{baida2009quantitative}). Dashed
line represents $\gamma _{R}((2R)^{-1})$ dependence (Equation (\ref{gammaR})
 for $A=0.7$ and $\gamma _{bulk}=0.125eV$). These
parameters \cite{baida2009quantitative} result from a linear fit to the
experimental data. Solid line results from our modeling in the extended
range of sizes for $\omega _{p}=9.10eV$, $\varepsilon
_{0}=4eV$, $\gamma _{bulk}=0.125eV$ and $A=0.7$. Our model with such input
parameters reproduces the experimental data for smallest particles perfectly
and describes the departure of $\omega _{l=1}((2R)^{-1})$ dependence from linear. The $\omega _{l}(R)$ model functions describe
consistently the size dependence of the SP damping rates for
experimentally available particles and predicts damping
rates (and the resulting damping times) for larger sizes (experimentally
unavailable so for) and for higher plasmon multipolarities.

Figure \ref{Fig_exp_gold}a) illustrates spectral widths $\Gamma$
vs resonance energy $\omega _{l=1}^{\prime }(R)$ measured
for different single Au nanoparticles (up to $2R=150nm$) (Figure 4 of \cite%
{SonnichenFranzlWilkVonPlessen2002}). The experimental data seems to suggest
that $\Gamma$ decays linearly with the SP resonance frequency.
However, it is not the case. The dependence of $\Gamma
_{l=1}(\omega_{l=1}(R))=2|\omega _{l=1}^{\prime \prime }(R)|$ (line in figure
\ref{Fig_exp_gold}a)) reproduces the experimental
data and predicts the decrease in $\Gamma_{l=1}$ for particles larger than those studied experimentally. Such decrease in damping rate $|\omega _{l=1}^{\prime \prime }(R)|$ of the dipole SP is accompanied by the
increase in the damping rate of the quadrupole $|\omega _{l=2}^{\prime
\prime }(R)|$ SP and of the following higher polarity SPs, as
discussed in Section \ref{Suppression} (see Figure \ref{Fig_eigenvalues_full}b)).

Figure \ref{Fig_exp_gold}b) illustrates the quality factor of the
dipole resonance $Q_{l=1}=\omega _{l=1}^{\prime }(R)/2|\omega _{l=1}^{\prime
\prime}(R)|$ vs resonance energy $\omega _{l=1}^{\prime }(R)$ (solid
lines). Our modeling shows that interface damping (dashed line)
substantially reduces the quality factor for particles from the smallest
size range. The maximum factor $Q_{l=1}\approx 26$ is found in nanoparticles of
radii $R=12nm$. The agreement with the experimental data of
\cite{SonnichenFranzlWilkVonPlessen2002} is very good.

\section{Conclusions}

The dependence of the SP resonance frequencies $\omega _{l}^{\prime }(R)$ and
damping rates $|\omega _{l}^{\prime \prime }(R)|$ on particle radius determine the intrinsic optical
properties of plasmonic spheres (illuminated or not). Our
study provides direct, accurate size characteristics for a broad range
of particle radii (up to $R=1000nm$) and plasmon polarities (up to $l=10$).
At present, this exceeds the range experimentally explored.

Size dependence of SP damping rates $|\omega _{l}^{\prime \prime }(R)|$ allows to distinguish the size ranges in which efficient transfer of radiation energy into heat takes place (large contribution of the nonradiative decay) and those in which the particles are effective radiating antennas (dominant contribution of the radiative damping). As long as the contribution of the radiative decay is negligible, the particle is not able to couple to the incoming field effectively and has weak radiative abilities. If the contribution of the radiative damping prevails, the particle is able to emit light within plasmonic mechanism efficiently. Effective radiating enables efficient interaction of SP near-field with other structures at a desired resonance frequency $\omega =\omega _{l}^{\prime }(R)$. The knowledge of size dependence of both: the multipolar SP resonance frequencies $\omega _{l}^{\prime }(R)$ and the corresponding damping rates $|\omega _{l}^{\prime \prime }(R)|=\hbar /T_{l}(R)$\ is indispensable to shape the particle plasmonic features effectively.

Our study, extended toward large particle sizes and plasmon multipolarities, revealed new features of the total plasmon damping rates. In certain ranges of radii, reduction of the multipolar SP damping rates, as compared with its small size limit, takes place. The small size limit is equal to the nonradiative damping rate resulting from absorption and heat dissipation. The suppression of nonradiative damping is not limited to the size range, in which the effect was directly demonstrated. It is present when the radiative damping brings the dominant contribution to the total plasmon damping. The reduction of $\gamma^{nrad}(R)$ with the growing contribution of radiation damping is revealed in the absorption spectra of particles; absorptive abilities of large particles are poor. This fact is described by solutions of Mie scattering theory, but its physical background have not been explained, as far as we know.
Our study led us to conclusion, that the reduction of multipolar SP nonradiative damping rates results from
competition between radiative damping and
all other damping processes included in $\gamma ^{nrad}(R)$. As far as we know, such hypothesis has not been proposed before. Our study provides a new starting point for better understanding of rules that define contributions of plasmon radiative and nonradiative decay rates to the total SP decay rate as a function of particle size. The independent modeling  is necessary for better understanding of the role of SP radiative and nonradiative decay channels.

The proposed SP characteristics vs particle size provide a consistent, uniform description of the experimental SP damping rates measured for single spherical particles in \cite{SonnichenFranzlWilkVonPlessen2002} and \cite{baida2009quantitative}. Not only dipole damping rates (found experimentally) but also multipole damping rates (and resulting damping times) in a broad range of sizes can be predicted.

The derived size dependence of plasmon decay rates $|\omega _{l}^{\prime
\prime}(R)|$\ at resonance frequencies $\omega _{l}^{\prime }(R$\ $)$, and the quality factors $Q_{l}(R)$\ of SP multipolar modes define not only the spectral scattering abilities of the plasmonic spheres, but also reflect changes in the strength of coupling of SP modes with the external field. Such
characteristics can serve as a tool for controlling the spectral features of
plasmonic nanospheres in technological or diagnostic applications, by
optimizing the size at the desired resonance frequency $\omega =\omega
_{l}^{\prime }(R)$.

\bigskip

\textbf{Acknowledgement} We would like to acknowledge the financial support
of this work by the Ministry of Science and Higher Education (No N N202
126837).

\bibliographystyle{elsarticle-num}
\bibliography{OpticalNanoantennas}

\end{document}